\documentclass[a4paper]{article}
\usepackage[left=2.5cm,top=2.5cm,right=2.5cm,bottom=2.5cm]{geometry}
\usepackage{fancyhdr}
\usepackage[utf8]{inputenc}
\usepackage[english]{babel}
\usepackage[T1]{fontenc}
\usepackage{siunitx}
\usepackage[nodayofweek]{datetime}
\usepackage{graphicx}
\usepackage[colorlinks=True]{hyperref}
\hypersetup{citecolor=blue}
\usepackage{amsmath}    
\usepackage{amssymb}    
\usepackage{amsbsy}    
\usepackage[square,numbers,semicolon]{natbib}
\usepackage{wrapfig}
\usepackage{enumitem}
\usepackage{multicol}
\newcommand{\noop}[1]{}

\title{\huge Data-driven spatio-temporal modelling of glioblastoma}
\date{}

\let\OLDthebibliography\thebibliography
\renewcommand\thebibliography[1]{
  \OLDthebibliography{#1}
  \setlength{\parskip}{0pt}
  \setlength{\itemsep}{0pt plus 0.3ex}
}

\usepackage{setspace}
\usepackage[table]{xcolor}

\begin{document}

\maketitle
\vspace{-1.5cm}
\noindent \large{Andreas Christ Sølvsten Jørgensen$^{1*}$, Ciaran Scott Hill$^{2,3}$, Marc Sturrock$^4$, Wenhao Tang$^1$, Saketh R. Karamched$^{5}$, Dunja Gorup$^{5}$, Mark F. Lythgoe$^{5}$, Simona Parrinello$^{3}$, Samuel Marguerat$^{6}$, Vahid Shahrezaei$^{1*}$} \\
\small{$^1$Department of Mathematics, Faculty of Natural Sciences, Imperial College London, London, UK \\
$^2$Department of Neurosurgery, The National Hospital for Neurology and Neurosurgery, London, UK \\
$^3$Samantha Dickson Brain Cancer Unit, UCL Cancer Institute, London WC1E 6DD, UK \\
$^4$Department of Physiology and Medical Physics, Royal College of Surgeons in Ireland, Dublin, Ireland \\
$^5$Division of Medicine, Centre for Advanced Biomedical Imaging, University College London (UCL), London, UK \\
$^6$Genomics Translational Technology Platform, UCL Cancer Institute, University College London, London, UK
}

\vspace{0.5 cm}

\noindent Correspondence*: 
Andreas Christ Sølvsten Jørgensen and Vahid Shahrezaei \\
a.joergensen@imperial.ac.uk, v.shahrezaei@imperial.ac.uk







\vspace{0.5 cm}

\noindent \textbf{Abstract:} Mathematical oncology provides unique and invaluable insights into tumour growth on both the microscopic and macroscopic levels. This review presents state-of-the-art modelling techniques and focuses on their role in understanding glioblastoma, a malignant form of brain cancer. For each approach, we summarise the scope, drawbacks, and assets. We highlight the potential clinical applications of each modelling technique and discuss the connections between the mathematical models and the molecular and imaging data used to inform them. By doing so, we aim to prime cancer researchers with current and emerging computational tools for understanding tumour progression. Finally, by providing an in-depth picture of the different modelling techniques, we also aim to assist researchers who seek to build and develop their own models and the associated inference frameworks.

\vspace{0.5 cm}

\section{Introduction}

Glioblastoma (GBM) is a malignant hierarchically organised brain cancer. It is both the most common and most aggressive type of primary brain cancer in adults \citep{Davis2016}. Not only does the diffusive invasion of glioma cancer cells into healthy tissue impede complete resection, but GBM harbours a subpopulation of highly therapy-resistant stem-like cells \citep{Lan2017}. Tumour recurrence is inevitable,  resulting in a median survival time of 15 months for patients despite maximal treatment \citep{Wang2015Dignosis}. The current gold standard of treatment is the Stupp protocol, which consists of maximal safe surgical resection, followed by radiotherapy and chemotherapy with temozolomide, an alkylating agent \citep{Stupp2005, Stupp2009}.

Mathematical cancer models have provided a deeper understanding of this immensely complex disease by unveiling the underlying mechanisms and offering quantitative insights \citep{Anderson2008, Altrock2015, Anderson2018, Cranmer2020}. Such models have entered all areas of GBM research, ranging from the classification and detection of brain tumours to therapy \citep{Zadeh2020, Finch2021}. 

This review provides the reader with an overview of existing mathematical and computational models that aim to simulate spatially resolved tumour growth. We discuss three main paradigms that have emerged for such \textit{in silico} experiments. In Section~\ref{sec:continuum}, we introduce so-called continuum models that treat variables, such as the tumour cell density, as continuous macroscopic quantities based on conservation laws. Alternatively, one might represent each cell as an individual agent. Such discrete models are discussed in Section~\ref{sec:discrete}. Section~\ref{sec:hybrid} deals with hybrid multi-scale and multi-resolution models that merge and bridge different approaches.

Each of the three methods has its advantages and shortcomings. One must choose between them based on computational limitations and the level of detail required to answer the research questions of interest. With this review, we aim to assist researchers in choosing between the different methods by highlighting the drawbacks and assets of each approach and by showing how the different methods can complement each other. Moreover, we summarise the main concepts and the key mathematical expressions that lie at the core of each approach. We hereby aim to strike a balance between providing a brief overview and showing the mathematics involved.

Mathematical models of GBM draw on a wide variety of molecular and imaging data: histopathological data, computerised tomography (CT), positron emission tomography (PET), single-photon emission computerised tomography (SPECT), and magnetic resonance imaging (MRI), such as T1 weighted (+/- Gadolinium contrast), T2 weighted, T2-FLAIR, and diffusion-weighted imaging (DWI), and most recently spatial and single cell transcriptomics. The models have thus been employed to shed light on patient-specific data \textit{in vivo} and \textit{ex vivo} as well as on data from animal models and \textit{in vitro} experiments. These analyses have provided invaluable insights and have deepened our understanding of glioma on a molecular and structural level. We address this issue in more detail in Section~\ref{sec:data}. The section also discusses the computational challenges posed by systematic inference of model parameters. Finally, Sections~\ref{sec:Clinical_app} and \ref{sec:disc} provides a short overview of some clinical applications of these models and an overall summary, respectively. 

While we discuss the different methods in light of GBM, we note that the same models are applied to other types of cancer. Indeed, the models build on concepts that are broadly used to study tissues. We do, therefore, not only cite sources that deal with GBM but occasionally refer the reader to illustrative papers from other areas of oncology and biology (see also \citep{Jarret2018, Karolak2018}). It is worth noting that the models are, in a broader sense, actually widely used across scientific disciplines. Throughout the paper, we thus present ideas and concepts that are also employed in other fields ranging from statistical mechanics to solid-state physics.
For instance, the cellular Potts model presented in Section~\ref{sec:CPM} builds on the so-called Ising model used to describe ferromagnetism. We hence encourage the reader to think outside the box when exploring the literature, and, in this spirit, we provide a few citations to areas outside the realm of biology.

Before commencing, we would like to point the reader towards other reviews and papers for further details. Lowengrub et al. \citep{Lowengrub2010} provide a detailed account of continuum models. For an elaborate discussion on discrete models, we refer the reader to Van Liedekerke et al. \citep{Liedekerke2015}. Both Metzcar et al. \citep{Metzcar2019} and Weeransinghe et al. \citep{Weerasinghe2019} give a brief overview of the topic and include a list of recent references. For insights into hybrid multi-scale modelling, we recommend Deisboeck et al. \citep{Deisboeck2010} and Chamseddine \& Rejniak \citep{Chamseddine2019}. Falco et al. \citep{jcm10102169} present a concise overview highlighting their clinical implications. Ellis et al. \citep{Ellis2019} focus on mathematical models that intratumour heterogeneity and tumour recurrence based on next-generation sequencing techniques. Alfonso et al. \citep{AlfonsoRev2017} highlight the challenges that mathematical models face when dealing with glioma invasion. Finally, for an overview of the biology of the GBM from a clinical perspective, we refer the reader to the recent reviews by Finch et al. and McKinnon et al. \citep{Finch2021, McKinnon2021}.


\section{Continuum models} \label{sec:continuum}

When dealing with cancer treatment, we face questions related to tumour size, shape and composition. These questions all address cancer on a \textit{macroscopic} scale. While macroscopic tumour dynamics emerge from interactions on a cellular level, it is possible to construct informative mathematical models without tracking individual cancer cells. Instead, the tumour and its environment can be represented as continuous variables that are governed by partial differential equations (PDE). Such continuum models capture many aspects of cancer \textit{in vitro}, \textit{in vivo}, and in patients. They can account for the impact of heterogeneous brain tissue on tumour growth, for different invasive tumour morphologies, and to some extent, even for potential tumour recurrence \citep[e.g.][]{ Frieboes2007, Swan2018}. Moreover, they have successfully been applied in studies on the impact of chemotherapy and repeated immuno-suppression treatment \citep{Swanson2002, Frieboes2006}. Continuum models have thus been employed in diagnosis and treatment planning based on patient-specific data \citep[e.g.][]{Swanson2000, Gu2012, Corwin2013, Neal2013, Jackson2015}.

However, PDEs smooth out small-scale fluctuations, which implies that continuum models do not apply to small cell populations, such as those found in the tumour margin. For such small cell numbers, stochastic events play a crucial role, and the applicability and predictive power of continuum models are limited. To properly understand cancer invasion, we must track individual cells. But to do so comes at a high or currently insurmountable computational cost (see Section~\ref{sec:discrete}). Thus the use of continuum models represents a trade-off that enables and supports scalability and mathematical insights. As a result, continuum models are widely used in the community \citep{Wang2009, Lowengrub2010, Weerasinghe2019}.

This section introduces the basic mathematical concepts of continuum models and their biological motivation. Any such model in the literature builds on reaction-diffusion equations, describing variables such as the tumour cell density, the tumour volume fraction, the nutrient (oxygen, glucose) concentration, the neovasculature, the enzyme concentration, or other properties of the extracellular matrix (ECM) \citep[e.g.][]{Zheng2005, Frieboes2006, Dattoli2014, powathil2012modeling, gerlee2012impact}. Considering any such variable, $\psi(\mathbf{x}, t)$, which is a function of position $\mathbf{x}$ and time $t$, we have for its rate of change with time  
\begin{equation}
\frac{\partial \psi}{\partial t} = -\nabla \mathbf{J} + S, \label{eq:massconv}
\end{equation}
where $\mathbf{J}$ is the flux of the considered variable, while $S$ is the sources and sinks for this variable. Thus, Equation~(\ref{eq:massconv}) constitutes a conservation law. The exact expression for $\mathbf{J}$ and $S$, as well as the boundary conditions, will depend on the variable in question. For instance, when dealing with the quasi-steady diffusion of nutrients, Equation~(\ref{eq:massconv}) generally takes the form \citep{Cristini2003, Sinek2004, Cristini2005, Mahlbacker2018, Bull2020MathematicalMR}
\begin{equation}
0 = D \nabla^2 n + S. \label{eq:nutrient}
\end{equation}
Here, $D$ denotes a diffusion coefficient, while $n(\mathbf{x})$ is the relevant nutrient concentration at the location $\mathbf{x}$ within the considered \textit{n}-dimensional domain, which is oftentimes denoted by $\Omega$. 

Alternatively, let's consider the (normalized) cancer cell density, $\rho(\mathbf{x},t)$, at a time $t$ and location $\mathbf{x}$. Many authors assume that the diffusion of cancer cells can be well-approximated by Fick's first law
\begin{equation}
\mathbf{J}= -D \nabla \rho, \label{eq:Fick1}
\end{equation}
where $D$ denotes a diffusion coefficient, which we discuss in detail in Section~\ref{sec:anidiff}. As regards the sources and sinks of the cancer cell density, it is commonly assumed that cell proliferation, i.e. tumour growth, is well-described by a logistic growth term \citep[e.g.][]{Suarez2012, Rutter2017}. So, Equation~(\ref{eq:massconv}) takes the form 
\begin{equation}
\frac{\partial \rho}{\partial t} = \nabla(D\nabla \rho) + \lambda\rho(1-\rho) \label{eq:Fickian}
\end{equation}
where $\lambda$ is the growth rate of the tumour cell population. Other authors assume exponential tumour growth, substituting the second term on the right-hand side by $\lambda \rho$ \cite[e.g.][]{Swanson2000}. Of course, more than one term might be necessary to summarise the relevant sources and sinks. Several more complex terms are, for instance, needed when considering differentiation between different interdependent tumour sub-populations (cf. Section~\ref{sec:lineage}).

It is worth stressing that Equation~(\ref{eq:massconv}) cannot stand on its own. Other relations and constraints, including (Neumann) boundary conditions, are needed. One might, for instance, naturally require that there is no flux at the boundary of the brain domain, i.e. that the tumour doesn't penetrate the patient's skull \citep[e.g.][]{Giatili2012, PainterHillen2013, Rutter2017}. This being said, Equation~(\ref{eq:massconv}) is the backbone of any continuum model that spatially resolves the tumour (see also Section~\ref{sec:ODE}).

While we focus on spatio-temporal cancer models in this review, it is also worth noting that Equation~(\ref{eq:massconv}) can be seen as a natural extension to non-spatial models for tumour growth. Indeed, if we drop any spatial dependence in Equation~(\ref{eq:massconv}), including the diffusion term, we end up with an ordinary differential equation (ODE) for the tumour cell density on the form
\begin{equation}
\frac{\mathrm{d} \rho}{\mathrm{d} t} = S. \label{eq:ns_ode}
\end{equation}
Depending on the source term, $S$, Equation~(\ref{eq:ns_ode}) might describe exponential, logistic or Gompertz tumour growth laws that serve as the foundation for non-spatial cancer models. Building on Equation~(\ref{eq:ns_ode}), one might thus construct a sophisticated network of coupled ODEs (or delay differential equations) that might differentiate between different tumour cell subpopulations or account for immune responses or the effect of cancer treatment \citep[e.g.][and references therein]{Eftimie2010, Murphy2016, Barros2021, Yu2021}.


\subsection{Anisotropic Diffusion} \label{sec:anidiff}

By using a scalar for the diffusion coefficient in Equation~(\ref{eq:Fickian}), we assume isotropic tumour growth. However, GBM spreads anisotropically, primarily expanding along pre-existing structures, such as blood vessels and white matter tracks \citep{Scherer333, Giese1996, Rao2003, Konukoglu2010, Gritsenko2012, Cuddapah2014, Brooks2020}. To take the heterogeneous structure of the brain into account, Swanson et al. \citep{Swanson2000} hence proposed to adopt different values for the diffusion coefficients in grey and white matter. Concretely, based on CT scans by Tracqui et al. \citep{Tracqui1995}, Swanson et al. \citep{Swanson2000} found the diffusion coefficient in white matter to be more than five times larger than in grey matter.

Taking the idea of heterogeneous diffusion further by including anisotropy, other authors \citep[e.g.][]{Jbabdi2005, Hogea2008, Konukoglu2010, PainterHillen2013, Antonopoulos2019} substitute the diffusion coefficient with an $n$-dimensional diffusion tensor, $D(\mathbf{x},t)\in \mathbb{R}^{n\times n}$. To evaluate $D(\mathbf{x},t)$, Painter \& Hillen \citep{PainterHillen2013} deploy diffusion tensor imaging (DTI) data. DTI is a magnetic resonance imaging (MRI) technique that measures the anisotropic diffusion of water molecules and hereby maps highly structured tissue. This technique provides the diffusion tensor for water molecules, $D_\Pi(\mathbf{x},t)$, throughout the brain. 
Of course, due to the size difference, the movement of cancer cells is more restricted than that of water molecules, which means that $D_\Pi(\mathbf{x},t)$ does not adequately describe glioma growth, i.e. $D(\mathbf{x},t)\neq D_\Pi(\mathbf{x},t)$. However, based on a transport equation for individual cell movement, Painter \& Hillen \citep{PainterHillen2013} establish a relation between $D(\mathbf{x},t)$ and $D_\Pi(\mathbf{x},t)$ expressed in terms of the Fractional Anisotropy (FA) that is commonly used to quantify DTI data \citep{Beppu2003}. They do so based on a set of simplifying assumptions and parabolic scaling to a macroscopic model \citep[see][for further details]{Hillen2013}. Their final macroscopic model takes the form
\begin{equation}
\frac{\partial \rho}{\partial t} = \nabla \nabla(D\rho) + \lambda \rho(1-\rho) \label{eq:Fokker},
\end{equation}
where the diffusion tensor $D \in \mathbb{R}^{n\times n}$ is symmetric and positive-definite, as it is related to the variance-covariance matrix of the probability distribution function that describes the velocity changes of individual cells \citep[see][]{Hillen2013, PainterHillen2013}. In other words, when dealing with three-dimension data, $D(\mathbf{x},t)$ is a symmetric and positive-definite $3\times 3$ matrix that incorporates the impact of the local environment on cell migration.
The model by Painter \& Hillen \citep{PainterHillen2013} has been extended and applied by other authors \citep{Engwer2015, Engwer2016, Swan2018}.

Note that Equation~(\ref{eq:Fokker}) is subtly different from Equation~(\ref{eq:Fickian}). Apart from $D$ denoting an $n\times n$ tensor rather than a scalar, Equation~(\ref{eq:Fokker}) includes an additional advective-type term since $\nabla \nabla (D\rho)= \nabla (D \nabla \rho) + \nabla (\nabla D \rho)$. Models of the form of Equation~(\ref{eq:Fokker}) are referred to as Fokker-Planck models, while Equation~(\ref{eq:Fickian}) is an example of a Fickian model. The additional advection term of the Fokker-Planck model has a demonstrable impact on the solution \citep[see also][]{Belmonte2013}. We also note that Equation~\ref{eq:Fokker} would correspond to the Fisher's equation if the first term on the right-hand side were substituted by $D\nabla^2\rho$. Indeed, some authors employ a diffusion term of this kind to describe the cancer cell density \citep{Lee2017}.


\subsection{Mechanical interactions, cell types, lineage, and feedback} \label{sec:lineage}

Tumours are hierarchically organised. GBMs harbour stem-like cells (GSC) as well as proliferating (GCP) and differentiating (GTP) subpopulations \citep[cf.][]{Singh2004, Venere2011, Lan2017, Brooks2020}. Since these three cell types exhibit very different behaviours, it is insightful to differentiate between these subpopulations, even when dealing with continuum models. 

Models that distinguish between viable and necrotic tumour tissue are the first step in this direction. Examples of such models can be found in the papers by Wise et al. \citep{Wise2008, Wise2011} and Frieboes et al. \citep{byrne1996growth, Frieboes2013}. Their work is based on reaction-diffusion equations of the form
\begin{equation}
\frac{\partial \rho_i}{\partial t} + \nabla (\mathbf{u}_i \rho_i) = - \nabla \mathbf{J}_{\mathrm{mec},i} + S_i \label{eq:Fickian2}
\end{equation}
where the index $i$ runs over all subpopulations, $\mathbf{u}_i$ denotes the velocity of the considered cell species, and $\mathbf{J}_{\mathrm{mec},i}$ is the flux that arises from mechanical interactions
\begin{equation}
\mathbf{J}_{\mathrm{mec},i} = \mathbf{J}_i - \rho_i \mathbf{u}_i. \label{eq:jmec}
\end{equation}

Note that while Equation~(\ref{eq:Fickian2}) appears to differ from the other reaction-diffusion equations listed above, this is merely a matter of notation. It's a Fickian model that can be derived by inserting Equations~(\ref{eq:Fick1}) and~(\ref{eq:jmec}) into  Equation~(\ref{eq:massconv}). The advantage of phrasing the problem in this manner is that the mechanical flux reflects the mechanical interaction energy that can be obtained from an understanding of the underlying cell biology \citep[see also][]{Kim2004a, Wise2004, Kim2005, Frieboes2007}. By introducing $\mathbf{J}_{\mathrm{mec},i}$, it is thus possible to inform the model about the properties of the tumour and host without the necessity of constructing a suitable diffusion tensor.

In papers that employ the Equation~(\ref{eq:jmec}), $\mathbf{u}_i$ is computed by imposing relations similar to Darcy's law that links the velocity to a gradient in pressure \citep[e.g.][]{Frieboes2007}. For glioma, the relevant expression often takes the form $\mathbf{u}_i=-\nabla p + \mathbf{F}_i$, where p is the solid pressure arising from the tumour proliferation, whilst $\mathbf{F}_i$ reflects mechanical interactions.

We exemplify the source functions that enter Equation~(\ref{eq:Fickian2}) by listing the relevant terms for the necrotic tissue according to Wise et al. \citep{Wise2008}, for which
\begin{equation}
S_\mathrm{d} = \lambda_\mathrm{A}\rho_\mathrm{v} + \lambda_\mathrm{N}\mathcal{H}(n_\mathrm{N}-n)\rho_\mathrm{v} - \lambda_\mathrm{C} \rho_\mathrm{d}.
\end{equation}
Here, the indices `d' and `v' refer to the dead and viable cancer cells, respectively, while $\lambda_\mathrm{A}$, $\lambda_\mathrm{N}$, and $\lambda_\mathrm{C}$ denote the rates of apoptosis, necrosis, and the clearance of dead cells. Moreover, $\mathcal{H}$ is a Heaviside step function, and $n_\mathrm{N}$ is a viability limit for the nutrient concentration below which cells die. For comparison, the source function for the viable tissue takes a similar form:
\begin{equation}
S_\mathrm{v} = -\lambda_\mathrm{A}\rho_\mathrm{v} - \lambda_\mathrm{N}\mathcal{H}(n_\mathrm{N}-n)\rho_\mathrm{v} + \lambda_\mathrm{M} \frac{n}{n_\infty} \rho_\mathrm{v},
\end{equation}
where $\lambda_\mathrm{M}$ denotes the rate of mitosis, and $n_\infty$ is the far-field nutrient level.

By distinguishing between GSCs, GCPs, and GTPs, Kunche et al. \citep{Kunche2016} and Yan et al. \citep{Yan2017, Yan2017_3D, Yan2018_3D} have incorporated the GBM lineage and hereby taken the discussed models one step further. Kunche et al. \citep{Kunche2016} use this to investigate feedback regulation of cell lineage progression. Furthermore, while previous papers only consider adhesion when computing the mechanical interactions, Chen et al. \citep{Chen2014a, Chen2014b, Chen2019} include the impact of elastic membranes and the implications of the calcification of dead tumour cells \citep{Panorchan2006}.


\subsection{Modelling the macroscopic environment} \label{sec:microenv}

Understanding the microenvironment is essential for understanding GBM since the brain region and other properties, such as the patient's age, have been shown to play a key role in tumour development and heterogeneity \citep[e.g.][]{Brooks2020, Ravi2021.02.16.431475}.

The concentrations of different chemicals, including (but not limited to) nutrients, drugs, matrix-degrading enzymes, and ECM macromolecules, are commonly modelled using reaction-diffusion equations. The associated diffusion is often in the form $D\nabla^2\psi$, but other second-order spatial derivatives can be found in the literature \citep{Anderson2006, Jiang2006, Powathil2012}. The source terms reflect the processes at play. For instance, the rate of oxygen consumption is often assumed to be proportional to the local oxygen concentration and might be proportional to the local cell density. 

Overall, the use of continuum models to model, say, nutrient flows is justified through the different scales that define cancer. Continuum models reliably capture the spatial gradients and temporal variation of nutrients in a way that allows us to assess the behaviour of tumour cells. For some purposes, it might even be adequate to assume that the nutrient levels stay constant over time since cell proliferation takes place on a much longer time scale than nutrient diffusion (cf. Equation~\ref{eq:nutrient}).

Initially, tumours do not possess their own vasculature but rather rely on the diffusion of nutrients and waste products. During this so-called avascular phase, the tumour may lack some features of malignancy and appear more benign. But beyond a certain size (1-3 mm in radius), the nutrient inflow can no longer sustain the growing cell population \citep{Mantzaris2004}. Hypoxia, i.e. oxygen shortage, sets in. Without stimulating angiogenesis, i.e. recruiting vasculature, tumour growth will stagnate. However, in response to the hypoxic conditions, the cancer cells release tumour angiogenic factors (TAFs), which stimulates the migration and proliferation of endothelial cells (ECs). New blood vessels sprout towards the tumour, and tumour growth resumes. Understanding the transition from avascular to vascular tumour growth is essential since it is a critical step toward malignancy. To study angiogenesis, many authors model both the diffusion of TAFS and ECs using reaction-diffusion equations \citep[e.g.][]{CHAPLAIN199647, Orme1997TwodimensionalMO,  Mantzaris2004, Lowengrub2010, Gu2012}. As regards the ECs, we again encounter logistic or exponential source terms. Meanwhile, the gradient of the TAF concentration, as well as that of adhesive molecules (e.g. fibronectin), enter through a flux term encapsulating chemotaxis and haptotaxis:
\begin{equation}
\mathbf{J} = -D \nabla \rho_{\mathrm{EC}} + \sum_i \chi_i \rho_{\mathrm{EC}}  \nabla c_i,
\end{equation}
where $\rho_{\mathrm{EC}}$ denotes the EC cell density, and the sum runs over all relevant diffusive components, including TAFs. The concentration of and sensitivity to the $i$th components are denoted as $c_i$ and  $\chi_i$, respectively \citep[see][]{Mantzaris2004}.

Like continuum models for tumour cells, continuum models for ECs cannot capture the behaviour of individual cells. Some authors, therefore, rely on agent-based models to study the ECs during angiogenesis \citep{Shirinifard2009, Daub2013, Xu2016, Chamseddine2018}. We refer to Section~\ref{sec:discrete} for a detailed discussion of agent-based models \citep[see also][]{Anderson1998}. Table~\ref{tab:CMs} summarises the properties of all continuum models discussed in this section.

\begin{table}
    \centering
    \caption{Overview of the reviewed continuum models. The references are not exhaustive.}
    \vspace{0.1cm}
    \label{tab:CMs}
    \rowcolors{2}{orange!10}{white}
    \begin{tabular}{|p{0.18\textwidth}|p{0.42\textwidth}|p{0.34\textwidth}|}
        \hline
        \rowcolor{orange!20} \textbf{Model type} & \textbf{Features} & \textbf{References} \\
        \hline
        Continuum model \newline (CM), tumour & \textbf{Concept:} The tumour cell density can be modelled as a fluid using PDEs based on conservation laws.
        \newline \textbf{Assets:} CMs capture macroscopic tumour growth at low computational cost.
        \newline \textbf{Drawbacks:} They cannot capture heterogeneity or the behaviour at low cell densities (e.g. at the tumour front).
        & \textbf{Methods:} Anisotropic diffusion tensor \citep{PainterHillen2013, Engwer2015, Swan2018} cf. Sec.~\ref{sec:anidiff}. Interaction energy-based models \citep{Wise2008, Frieboes2007, Yan2018_3D, Chen2019} accounting for lineage \citep{Yan2017} cf. Sec.~\ref{sec:lineage}. 
        \textbf{Patient-specific studies:} \citep{Gu2012, Neal2013, Jackson2015} \\ [2pt]
        CM, microenvironment & \textbf{Concept:} Different properties of the environment, such as nutrient flow or vasculature, are modelled on a macroscopic scale. 
        \newline \textbf{Assets:} They are computed at a low computational cost. Spatial gradients and time variations, e.g. in nutrient levels, are sufficiently well resolved to study tumour cells.
        \newline \textbf{Drawbacks:} CMs of ECs do not capture the behaviour of individual cells.
        & 
        \textbf{Nutrient flow:} \citep{Anderson2006, Jiang2006, Powathil2012}.
        \newline \textbf{Vasculature:} \citep{Mantzaris2004, Lowengrub2010}
        \newline These models are widely used whether the tumour is represented by a CM or by agent-based models (cf. Sec.~\ref{sec:discrete} and \ref{sec:hybrid}).
        \\ [2pt]
        \hline
\end{tabular}
\end{table}


\section{Discrete (agent-based) models} \label{sec:discrete}

Continuum models are based on the assumption that the behaviour of tumour cells is well-approximated by macroscopic averages. However, this assumption breaks down for small cell populations since stochastic events dominate these. Therefore, continuum models cannot reliably capture the onset of tumour growth, and they do not give a complete picture of the invasive tumour front or margin, i.e. the tumour-host interface, where the cancer cell density is low --- although it is possible to address this drawback by linking different scales, as shown by Trucu et al. \citep{Trucu2013} (see also Section~\ref{sec:hybrid}). Due to the aggressive invasion of healthy tissue by GBM, this shortcoming is especially problematic for understanding brain tumours. Moreover, continuum models cannot adequately deal with heterogeneity. To address these issues, one needs to model cancer on a mesoscopic scale following individual tumour cells or cell clusters \citep[see also][]{Hanahan2011}. Mathematical models that do so are referred to by various names: discrete models, agent-based models, individual-based models, or cell-based models. In all cases, cells or cell clusters are modelled as autonomous agents that follow a set of (stochastic) rules prescribing cell movement, death, division, and growth. By doing so, discrete models capture the variability among cells that might, for instance, arise as a response to the microenvironment \citep[e.g.][]{Anderson2006}. Note that while discrete models treat cells as autonomous quantities, continuum models are commonly used to depict the microenvironment (cf. Section~\ref{sec:composite}). PDEs hence still come into play when dealing with, e.g. nutrient and drug concentrations \citep[e.g.][]{Lowengrub2010, Macklin2016}.

Discrete models that describe spatial tumour growth or other cell populations, such as ECs (cf. Section~\ref{sec:microenv}), fall into two categories: lattice-based and off-lattice methods. While the former operates on a fixed lattice, the latter does not impose the same restrictions. Lattice-based models can be divided into three sub-categories \citep{Liedekerke2015, Metzcar2019}. First, there exists a group of so-called cellular automata (CA) models that treat all processes, including cell movement, as stochastic processes (cf. Section~\ref{sec:CA}). For these models, each lattice site (e.g. pixel or voxel) can either at most be occupied by a single cell or a population of a limited size \citep{Kansal2000b, Kansal2000, Mansury2002, Schmitz2002}. Each cell or cell cluster is thus characterised by its position on the grid. Secondly, there exist so-called lattice gas cellular automata (LGCA). While these models consider single cells and are conceptually very similar, they differ from other CA models by attributing a velocity to each cell (cf. Section~\ref{sec:LGCA}). After accounting for stochastic events, these models thus include a deterministic evaluation of cell propagation at every time step. Thirdly, there are so-called cellular Potts models (CPM) that phrase cell interactions within the framework of statistical mechanics (cf. Section~\ref{sec:CPM}). CPMs attribute several lattice points to each cell, whereby these models capture cell morphologies. Off-lattice models, also known as lattice-free models, come in two flavours. Centre-based models (CBM) describe cells as simple particles whose interactions can be expressed as physical forces. Deformable cell models (DCM) or vertex models extend this picture by resolving the cells using several nodes. Like CPMs, DCMs can thereby account for cell morphologies.

One drawback of discrete models is their relatively high computational expense due to the large number of cells that typically need to be simulated. After all, tumours reach $10^5-10^6$ cells per mm$^3$. However, one benefit of discrete models is their ability to capture emergent properties. The simple rules that govern the (nonlinear) interactions between individual agents will naturally give rise to the dynamics of macroscopic tumour growth. As discussed in Section~\ref{sec:LGCA}, reaction-diffusion equations might thus naturally follow from up-scaling agent-based models \citep{Hatzikirou2010, Hatzikirou2012, Jagiella2016}. Due to the ability of discrete models to capture emergent properties, agent-based models can be calibrated based on patient-specific data and be used to predict macroscopic parameters \citep{Macklin2012, Hyun2013}.

Concerning the implementation of discrete models, it is worth mentioning that some software tools are publicly available. There is no need for a research group to start from scratch. For a recent overview of open-source toolkits, we kindly refer the reader to Table 1 in the paper by Metzcar et al. \citep{Metzcar2019}. In this connection, we note that one must always select and adapt the numerical approach based on the problem at hand. For instance, not all problems require the same geometry. While colon cancer cells and cells \textit{in vitro} move in two dimensions, glioblastoma tumour growth \textit{in vivo} is intrinsically three-dimensional, which affects the growth pattern \citep{Ghaffarizadeh2018}. In the following two sections, we will discuss lattice-based models in more depth (Section~\ref{sec:latticemodels}) and then turn to a detailed discussion of off-lattice models (Section~\ref{sec:lattice-free}).


\subsection{Lattice-based models} \label{sec:latticemodels}

In lattice-based models, we follow a population of cells on a rigid grid. If a regular (e.g. Cartesian) lattice is chosen, artefacts in the established tumour growth pattern might arise, reflecting the lattice symmetries \citep{Liedekerke2015}. While unstructured meshes are harder to implement and combine with existing PDE solvers, they do not suffer from this shortcoming \citep{Block2008}.


\subsubsection{Cellular automata with one or several cells per lattice site} \label{sec:CA}

In this section, we discuss two types of cellular automata (CA) models: those with at most one and those with multiple cells per lattice site. Both approaches have their advantages and drawbacks. Below, we will start by discussing the single-cell CA models and then continue to frameworks that include many cells per lattice site.

When considering single tumour cells, we follow each of these through a sequence of time steps. During every time-step ($t\rightarrow t+\Delta t)$, a cell might migrate to a neighbouring lattice site, it might die, it might divide, and it might grow, whereby it temporally occupies two sites before dividing. Each of these events can be modelled as stochastic processes or be imposed through deterministic conditional statements that encompass our knowledge about cancer --- additional parameters, such as age, mutation, or phenotype, might be included in these rules \citep{Poleszczuk2016}. For instance, while cells might die spontaneously, external factors can make cell death increasingly likely. Thus, tumours develop a necrotic core due to nutrient deprivation. One might, therefore, set the probability for individual cell death to depend on the distance to the tumour front \citep{Schmitz2002}, on the local nutrient concentration, or the local level of toxic metabolites \citep{Mansury2002}. 

Tumour cell migration is also affected by the location of the cell within the tumour. If we consider a cell that is deeply buried within the tumour, all surrounding lattice sites are already occupied, and there is nowhere to migrate. This notion also plays a role in cell proliferation. To divide, a cell deep within the tumour must push its neighbours away to make space for the daughter cell,
which becomes increasingly difficult as the distance to the tumour edge increases. For simplicity, many authors account for this notion by introducing a sharp cut-off: If there are no free sites within a certain distance, the cell will be unable to divide \citep{Drasdo2005b}. Other models assume a smooth decline in proliferation with depth \citep{Jagiella2016, Jagiella2017}. On a macroscopic level, this suppression of cell division leads to a constant proliferative rim, below which we encounter a quiescent cell population surrounding a hypoxic region with a necrotic core at its centre. In accordance with data, this model predicts that a tumour spheroid enters linear growth after initially growing exponentially \citep{Hoehme2010}. As regards cell division, we also note that experiments suggest that the cell cycle time follows a $\Gamma$-like distribution, e.g. an Erlang distribution \citep{Radszuweit2009}. The rules imposed on the individual cells can account for a broad range of cell properties. For instance, if the CA model accounts for the hierarchical nature of tumour cells (GSCs, GCPs and GTPs), the probability of cell proliferation will also depend on the attributed cell-type \citep{Poleszczuk2016}.

When solving the system numerically, one must choose a suitable time-step, $\Delta t$. For this purpose, it is helpful to consider the cell population as a whole. Every time a cell dies, moves, divides, or grows, the system as a whole will transition into a new state. The time-step should be chosen accordingly such that only one event is likely to occur within $\Delta t$, and thus $\Delta t$ is itself a function of time. To achieve this, one might distil the temporal dynamics of the system into a master equation and employ kinetic Monte Carlo algorithms \citep{Gillesple1977, Bonabeau2002, Jagiella2016}. Moreover, the order in which the individual cells are updated should be random to avoid grid artefacts \citep{Lowengrub2010}.

Rather than considering individual cells, one might track cell clusters of a limited size, keeping track of the number of cells at each lattice site. The underlying concepts stay the same. The advantage of reducing the resolution in this manner is the reduced computational cost, which allows for modelling tumours of centimetre size \citep{Radszuweit2009}. Meanwhile, a coarser resolution might lead to artefacts affecting the predicted tumour growth. Another drawback of both single- and multiple-cell CA models is the fact that they do not properly account for cell morphologies. For models that address this issue, see Section~\ref{sec:CPM}.


\subsubsection{Lattice gas cellular automata models} \label{sec:LGCA}

Like the models in Section~\ref{sec:CA}, lattice gas cellular automata (LGCA) models follow a set of stochastic rules to determine whether individual cells die, grow, or divide. But in addition to specifying the location of the cells on the lattice, LGCA models attribute a discrete set of velocity channels to each spatial location. Hence, the cells roam a discretised phase space. 

Analogously to migration in the CA models presented above, cells might stochastically migrate in phase-space: They can move to neighbouring \textit{velocity channels} following probabilistic rules. At most, one cell can occupy any given velocity channel at a given location. This restriction is not to say that these channels are unique, i.e. at each location, there might be one or more channels with the same velocity, including channels at which the cell is at rest.

\begin{figure}
\centering
\includegraphics[width=0.75\linewidth]{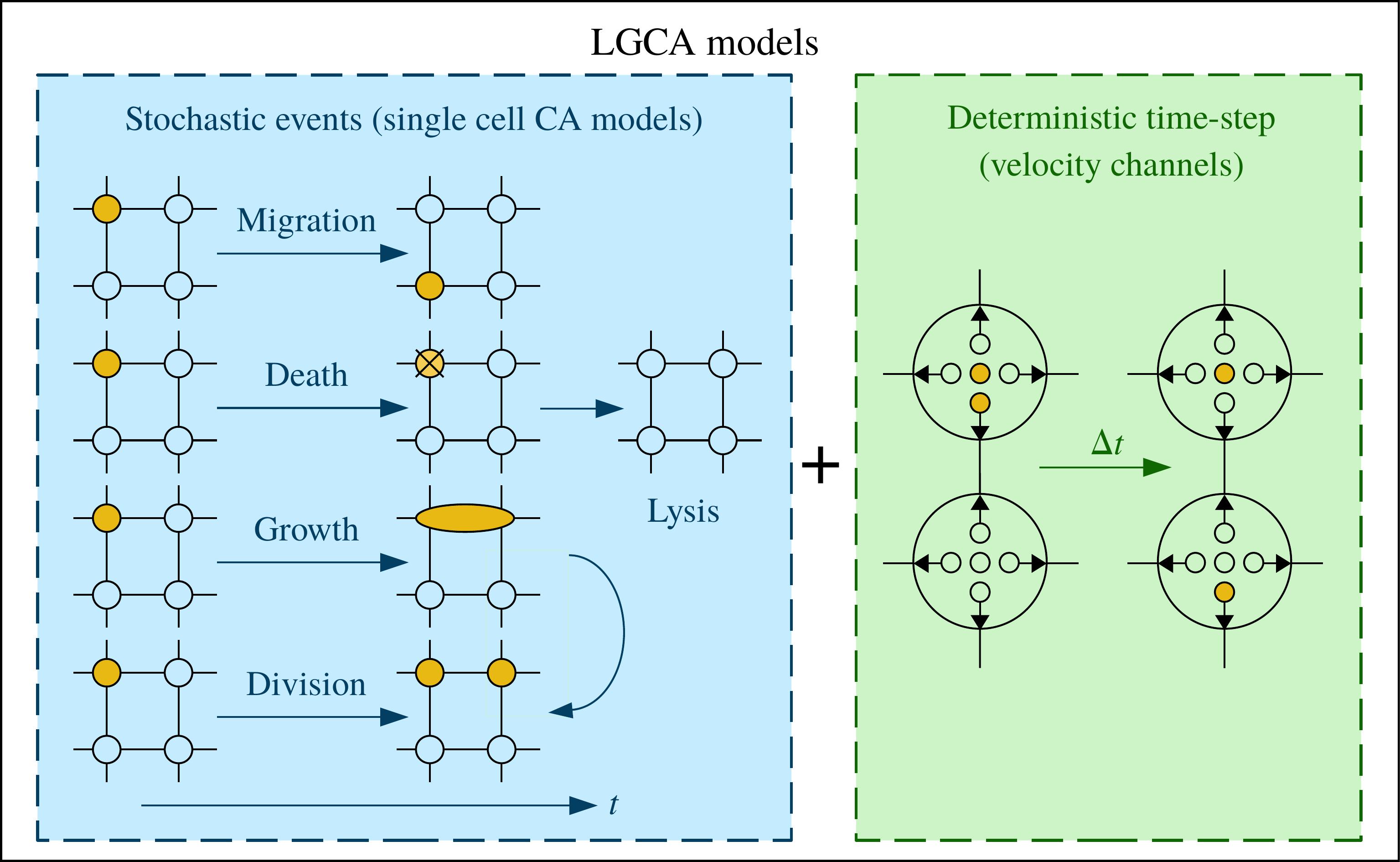}
\caption{Schematic overview of the concepts behind lattice gas cellular automata (LGCA) models. The \textbf{left panel} summarises the cell dynamics that are handled as stochastic processes: cell migration, death (and lysis), growth, and cell division. Filled (orange) circles denote occupied states, while there are no cells in the empty circles. On its own, the left panel corresponds to the cellular automata models with at most one cell per lattice site that are discussed in Section~\ref{sec:CA}. However, as illustrated in the \textbf{right panel}, several velocity channels are associated with each location, and multiple cells can occupy the same position. Filled orange circles denote occupied channels, while empty channels are indicated with empty circles. Here, there are five velocity channels per location: four channels that lead to migration and one channel at which the cell is at rest. After determining the stochastic processes, the model computes the movement of the cells based on the velocity channels that they occupy.}
\label{fig:LGCA}
\end{figure}

When using LGCA models, the time evolution of the system is split into two distinct steps. First, the model allows for a number of stochastic events that correspond to the number of cells. Afterwards, the model updates the position of every cell based on their individual velocities complying with momentum and mass conservation. In this manner, the model alternates between accounting for probabilistic interactions and deterministic cell movement. These basic concepts of LGCA models are summarised in Fig.~\ref{fig:LGCA}.

Within the mean-field approximation, Hatzikirou \& Deutsch \citep{Hatzikirou2010} show that the Boltzmann equation governs the macroscopic dynamics of LGCA models (The Boltzmann equation can be used to describe the evolution of thermodynamic systems that are out of equilibrium. Applications range from quantum mechanics over cosmology to magnetic resonance imaging \citep[e.g.][]{Dodelson2003}). Furthermore, through Taylor expansions and scaling, they arrive at a reaction-diffusion equation with a logistic growth term, giving a microscopic justification for the success of the models presented in Section~\ref{sec:continuum}.

LGCA models offer an intuitive implementation of cell migration. By comparing model predictions to experimental data, the group of Deutsch and Hatzikirou has thus used such models to gain insights into cell-cell interactions and mechanisms that underlie tumour invasion \citep{Chopard2010, Tektonidis2011, Hatzikirou2012}. In particular, they focus on the so-called "Go or Grow" (GoG) hypothesis. This hypothesis addresses the fact that glioma cells exhibit an inverse correlation between cell migration and proliferation \citep{Giese2003}. It states that migration and proliferation are mutually exclusive events, i.e. that moving cells cannot divide. This dichotomy stems from shared signalling pathways \citep[see also][]{Godlewski2010}. Since the transition from benign to malignant tumour growth is coupled with a transition from a highly proliferative to a highly migrative phenotype, it is paramount to understand the link between these two phenomena. Specifically, the group of Deutsch and Hatzikirou investigate the role of hypoxia \citep[see also][]{Monteiro2017}.


\subsubsection{Cellular Potts models} \label{sec:CPM}

\begin{figure}
\centering
\includegraphics[width=0.40\linewidth]{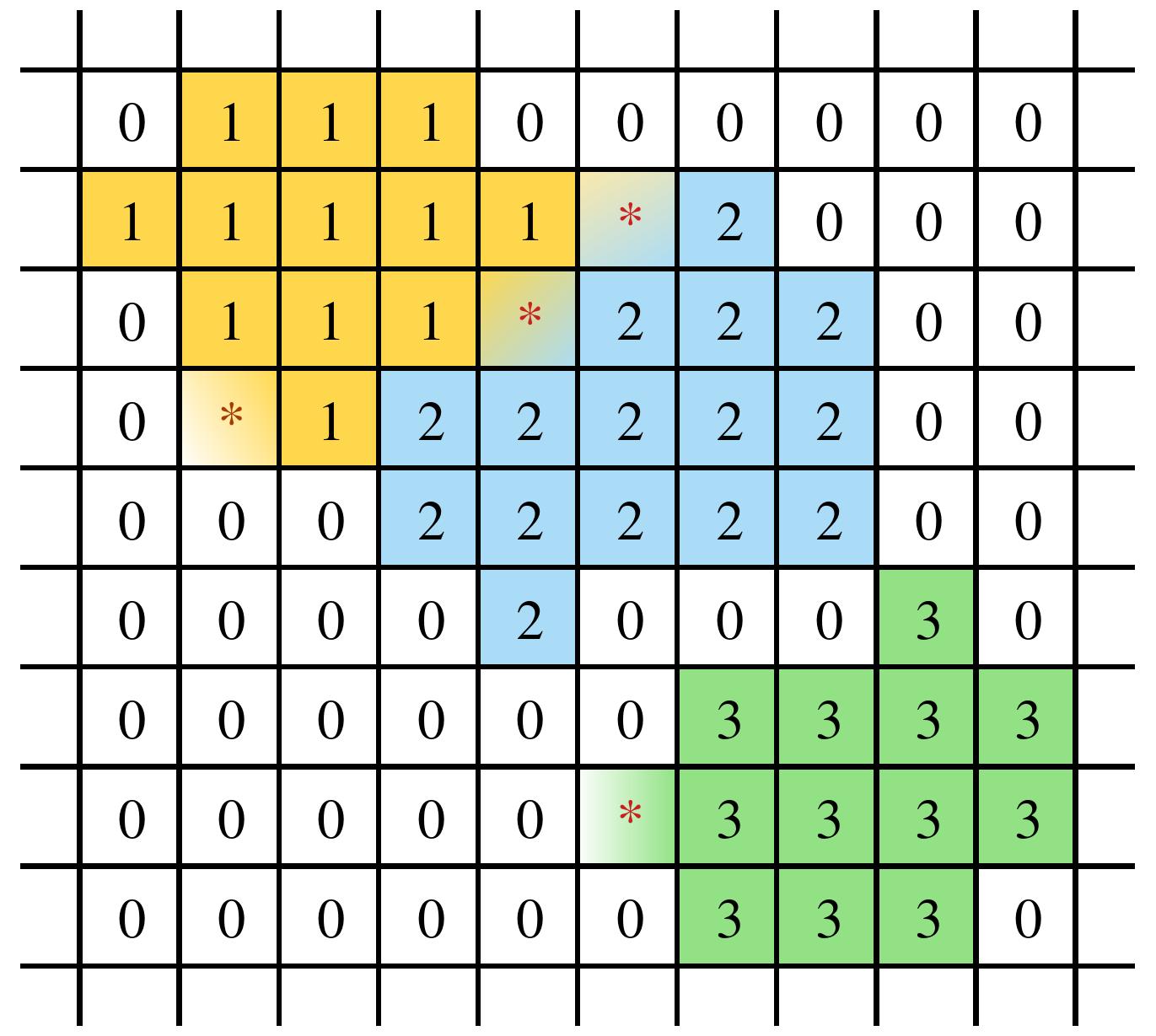}
\caption{Schematic overview of cellular Potts models. Each cell (and the ECM) has a unique identifier. Here, we consider cells 1, 2, and 3 and the ECM denoted by 0. Each cell occupies several lattice sites. Randomly selected sites at the cell borders (*) might swap affiliation depending on the energy change associated with this swap allowing the cells to grow and move.}
\label{fig:Potts}
\end{figure}

Cellular Potts models (CPM) define a Hamiltonian function, $H$, to incorporate cell movement, growth, interactions between adjacent cells, and interactions between cells and their microenvironment \citep[see][for a detailed discussion of Hamiltonian mechanics ]{taylor2005classical}. Moreover, CPMs attribute multiple lattice sites to each cell, which allows the models to account for cell morphology. Again, CPMs do so via the Hamiltonian function. The basic concepts are illustrated in Fig.~\ref{fig:Potts}

For simplicity, let's consider a Hamiltonian of the form \citep{Turner2002, Shirinifard2009}
\begin{equation}
H = \sum_{( \textbf{x,x'} )}E(\tau(\sigma(\textbf{x})), \tau(\sigma(\textbf{x'}))) \left( 1- \delta\left( \sigma(\textbf{x}), \sigma(\textbf{x'}) \right) \right) + \lambda_\mathrm{V} \sum_\sigma (V(\sigma)-V_\mathrm{T})^2. \label{eq:Hamil}
\end{equation}
The first term on the right-hand side of Equation~(\ref{eq:Hamil}) summarises the contact energy at the cell interfaces. The sum runs over all pairs of adjacent lattice sites $(\textbf{x,x'})$. Each cell (and the ECM) has a unique identifier $\sigma$, i.e. lattice sites with the same value of $\sigma$ are associated with the same cell (or the ECM). Moreover, each cell has a type $\tau$ \citep[e.g. GSC, GCP, or GTP, see][]{Gao2012}. $E(\tau(\sigma(\textbf{x})), \tau(\sigma(\textbf{x'})))$ denotes the contact energy per unit surface area (in 3D) between a cell of type $\tau(\sigma(\textbf{x}))$ and a cell of type $\tau(\sigma(\textbf{x'}))$. The Kronecker delta, $\delta\left( \sigma(\textbf{x}), \sigma(\textbf{x'})\right)$, is needed since we only encounter an interface if $\sigma(\textbf{x})\neq \sigma(\textbf{x'})$, while there will be an contact energy associated with cells of the same type. The second term on the right-hand side of Equation~(\ref{eq:Hamil}) represents a penalty for deviations between the volume, $V(\sigma)$, of a cell and its target size, $V_\mathrm{T}$. Here, the sum runs over all cells.

At every time-step, the CPM selects a random set of lattice sites at the cell boundaries and attempts to overwrite their cell affiliations ($\sigma$) with those of neighbouring sites: $\sigma(\textbf{x}) \rightarrow \sigma(\textbf{x'})$. In this manner, cells grow and move by high-jacking adjacent sites, whereby they might have to push other cells aside. Whether or not these proliferation and growth attempts are successful is decided by a Metropolis algorithm based on the associated change in the Hamiltonian function \citep[see][for a discussion of such algorithms]{Gregory2005}. The proposed change in cell affiliation is accepted with the probability
\begin{equation}
P(\sigma(\textbf{x}) \rightarrow \sigma(\textbf{x'})) = \min \left(1,\exp\left(-\frac{\Delta H}{\beta}\right)\right),
\end{equation}
where the Boltzmann temperature, $\beta$, describes the motility of the cells.

One can include additional terms in Equation~(\ref{eq:Hamil}) to add more information about cell biology. For instance, in analogy to the volume constraint, Ouchi et al. \citep{Ouchi2003ImprovingTR} suggest including constraints on the surface area of the cell (in 3D) \citep[see also][]{Boghart2014}. Other authors include chemotaxis, motility, haptotaxis, and haptokinesis in this manner \citep{Savill1997, Daub2013, Szabo2013, Li2014}. One immediate drawback of CPMs is that they are limited to phenomena and cell properties that can be formulated as terms in a Hamiltonian.

Meanwhile, when accounted for, cell division and death are treated as stochastic events or occur when certain conditions are met. For instance, Gao et al. \citep{Gao2012} let cells attempt division when $V(\sigma)>2V_\mathrm{T}$ and impose apoptosis for proliferating glioma cells after a pre-determined number of division attempts. Other authors impose (probabilistic) conditions on apoptosis and necrosis based on nutrient levels or overcrowding \citep{Shirinifard2009, Boghart2014}.


\subsection{Off-lattice models} \label{sec:lattice-free}

\begin{figure}
\centering
\includegraphics[width=0.75\linewidth]{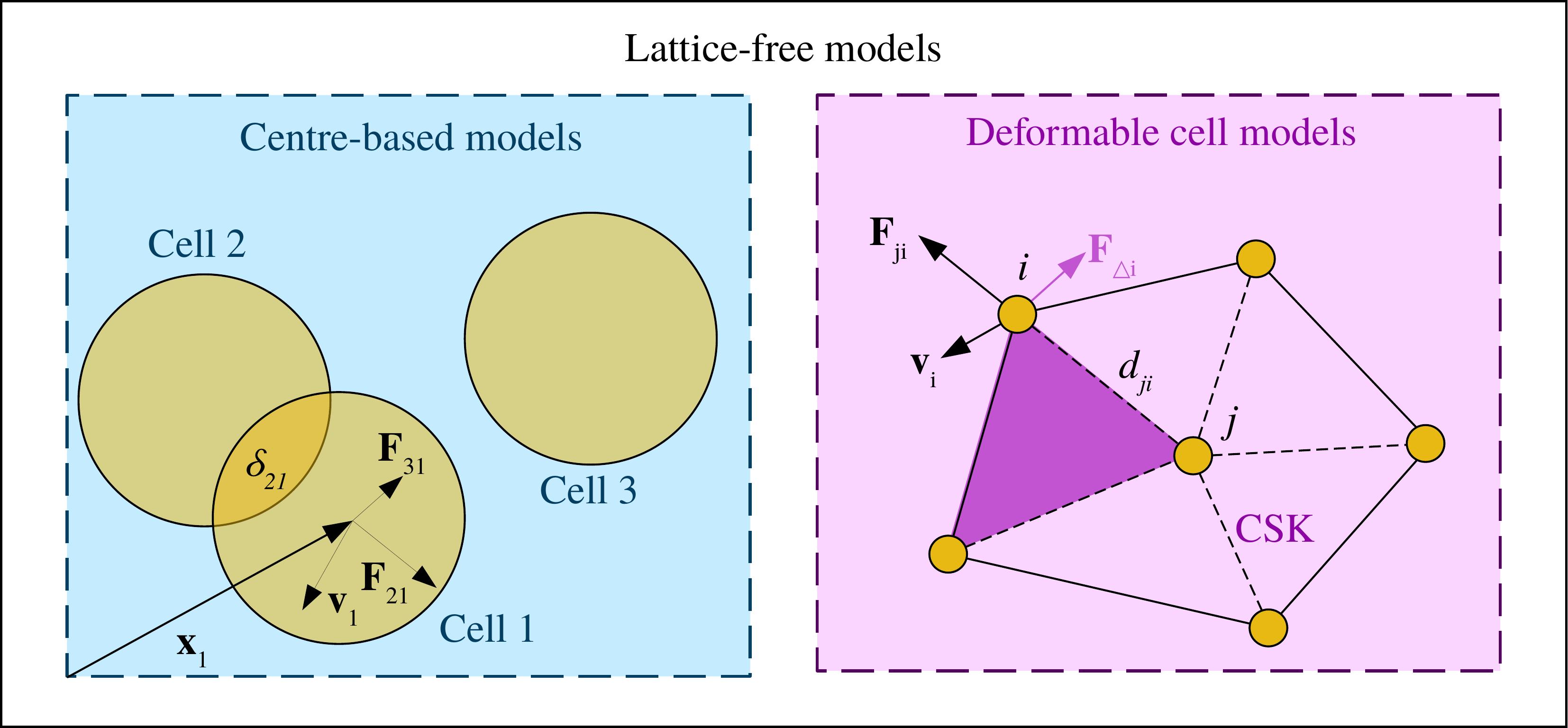}
\caption{Schematic overview of off-lattice models. The \textbf{left panel} illustrates spherical cells in a centre-based model (CBM), highlighting forces on cell 1. Due to the overlap ($\delta_{21}$) between cell 1 and 2, cell 1 is repulsed by cell 2. Overlapping cells forming a dumbbell also occur during cell division. $\mathbf{F}_{31}$ exemplifies an adhesive force occurring without direct contact. In an isotropic environment, the drag force will be proportional to the velocity, $\mathbf{v}_1$. The \textbf{right panel} illustrates a cell made of multiple nodes in a deformable cell model (DCM). Node $i$ at the cell membrane is subject to forces from the surrounding nodes ($\mathbf{F}_{ji}$). Constraints on areas and volumes might lead to additional forces ($\mathbf{F}_{\bigtriangleup i}$), while the movement of the node ($\mathbf{v}_i$) gives rise to viscous forces. The cell has a cytoskeleton (CSK, dashed lines). 
}
\label{fig:DCM}
\end{figure}

One of the main drawbacks of lattice-based models is that the lattice introduces a lower length scale. Off-lattice models overcome this problem at the expense of increased computational cost. We give a schematic overview of the two different types of off-lattice models in Fig.~\ref{fig:DCM}.

\subsubsection{Center-based models} \label{sec:CMs}

Centre-based models (CBMs) treat cancer cells as physical particles whose trajectories can be deduced from their equations of motion \citep[see][]{taylor2005classical}. The cells themselves are usually represented as spheres (in 3D) or viscoelastic ellipsoids that deform when subjected to external forces \citep{Dallon2004HowCM}. During mitosis, the mother cells are oftentimes modelled as dumbbells, consisting of two overlapping spherical cells \citep{Drasdo2005}. 

The equations of motion can be derived within an energy-based picture or by modelling the relevant interactions as physical forces that act on the particles. In the latter approach, the models draw on Newton's second law and commonly assume that inertial forces and any average acceleration can be neglected (cf. Brownian dynamics). Put another way, we are dealing with friction-dominated overdamped motion, which means that friction forces balance out all the other forces \citep{Drasdo2005b, Hoehme2010, Liedekerke2015}. For cell $i$ on a substrate, we can then write \citep[cf.][]{Liedekerke2015}
\begin{equation}
\Gamma_i^\mathrm{cs}\frac{\mathrm{d}\mathbf{x}_i}{\mathrm{d}t} + \sum_{j\neq i} \Gamma_{ji}^\mathrm{cc} \left( \frac{\mathrm{d}\mathbf{x}_i}{\mathrm{d}t} - \frac{\mathrm{d}\mathbf{x}_j}{\mathrm{d}t} \right) = \mathbf{F}_i^\mathrm{sub} + \mathbf{F}_i^\mathrm{mig} + \sum_j (\mathbf{F}_{ji}^\mathrm{adh} + \mathbf{F}_{ji}^\mathrm{rep}) \label{eq:eqmo}
\end{equation}
The first term on the left-hand side expresses cell-substrate friction forces, while the second term deals with friction between cells. Here, $\Gamma_{ji}^\mathrm{cs}$ and $\Gamma_{ji}^\mathrm{cc}$ are tensors that account for anisotropies and inhomogeneities. For instance, for spherical cells in a homogeneous and isotropic environment, $\Gamma_{ji}^\mathrm{cs}=\gamma \mathbb{I}$, where $\gamma$ is a damping coefficient, and $\mathbb{I}$ is the identity matrix \citep{Drasdo2005b}. Thus, the drag force that acts on a cell in an isotropic viscous environment is proportional to the cells' velocity \citep{Mathias2020}.

On the right-hand side of Equation~(\ref{eq:eqmo}), $F_i^\mathrm{sub}$ denotes forces that emerge from adhesion and repulsion between cell $i$ and the substrate. $F_i^\mathrm{mig}$ describes the migration forces. For instance, chemotaxis might introduce a preferred direction for migration \citep{Drasdo2005b}. One might account for the memory of cells, i.e. persistence of a cell to move in a given direction. Some authors do so using inertial forces. In addition, some authors include cell polarity giving the cell a preferred direction of motion. $F_i^\mathrm{mig}$ becomes a noise term that describes a random movement in the absence of such forces and assumptions. The last term on the right-hand side summarises the adhesion and repulsion between cell $i$ and all other cells. These forces will typically depend upon the distance between the cell centres or the amount of overlap, $\delta_{ji}$. Contact mechanics offers different approaches to model these cell-cell interactions, including Hertzian formulations and the Johnson-Kendall-Roberts (JKR) model of elastic contact \citep[cf.][]{Drasdo2005, Hatzikirou2010}. Some authors simply assume that all forces involved in cell-cell interactions are well-approximated by linear springs, generalised linear springs, or polynomial expressions \citep[e.g.][and references therein]{Meineke2001CellMA, Bull2020MathematicalMR, Mathias2020}.

To exemplify the use of Equation~(\ref{eq:eqmo}), let's consider spherical cells, whose movement is dominated by isotropic friction on the substrate and cell-cell interactions. Meineke et al. \citep{Meineke2001CellMA} and Drasdo \citep{Drasdo2005b} use this model to describe cellular mono-layers in the intestinal crypt and \textit{in vitro}. For mono-layers, we might assume that all cells have roughly the same velocity, which leaves us with the first term on the left-hand side and the last term on the right-hand side of Equation~(\ref{eq:eqmo}):
\begin{equation}
\gamma \frac{\mathrm{d}\mathbf{x}_i}{\mathrm{d}t} = \sum_j (F_{ji}^\mathrm{adh}+F_{ji}^\mathrm{rep}), \label{eq:simpleCPM}
\end{equation}
which amounts to a set of coupled ODEs that can be solved using explicit or implicit schemes, such as a forward Euler scheme \citep{Meineke2001CellMA, Mathias2020}. In general, Equation~(\ref{eq:eqmo}) can be written as a set of linear equations and be solved using matrix manipulation, for which standard software packages are available. 

The computational challenge of CBMs is two-fold: First, all interacting cell pairs must be identified. Secondly, the relevant interactions take place on chemical and dynamic time scales that are much shorter than the time scale that governs tumour growths. Besides their high computational cost, it should be noted that CBMs might suffer from artefacts when cells are densely packed \citep{Liedekerke2015}. Despite these shortcomings, CBMs have widely and successfully been applied to study tumour-related issues, such as drug response \citep[e.g.][]{Gallaher2020, Perez-Velazquez2020}.


\subsubsection{Deformable cell (vertex) models} \label{sec:DMs}

Deformable cell models (DCMs) are conceptually similar to CBMs: Cell dynamics are distilled into equations of motion. However, rather than representing each cell by simple geometrical objects, DCMs depict cells as a number of connected nodes (cf. Fig.~\ref{fig:DCM}). In contrast to CBMs, DCMs hereby capture cell morphology and intracellular mechanics. The nodes within a cell are connected by viscoelastic elements that are either part of the cell membrane or the cytoskeleton. It is essential to include a cytoskeleton since the modelled cells might buckle without an internal scaffolding structure \citep{Buenemann2008}.

Assuming that inertial forces can be neglected, the equation of motion for any given node takes the same form as that of a single cell in CBMs (cf. Equation~\ref{eq:eqmo}). Thus, frictional forces balance out all other forces acting on the node \citep[see also][]{Liedekerke2019b}. Node-node interactions can be modelled by letting the viscoelastic elements act as damped linear springs. To better depict the behaviour of biopolymers, one might model the elements using nonlinear forces \citep{Liedekerke2010, Odenthal2013}. Besides node-node interactions, we must introduce forces that express area and volume constraints on the cell as a whole or on subdivisions spanned by several nodes. By considering subdivisions of the cell membrane, we might furthermore introduce forces that express the resistance of the cortex to bend \citep{Odenthal2013}.

Simulating realistic cell dynamics requires a large number of nodes per cell, and the models can become rather complex. Due to the resulting high computational cost, it is often only feasible to simulate single cells or small cell populations on short time scales.

For completeness, it should be mentioned that there is a subgroup of DCMs called vertex models (VMs). In such models, the cells are represented as polygons spanned by a grid of nodes, i.e. adjacent cells share vertices and edges. There is no space between cells. VMs are useful when dealing with densely packed cells and have thus found applications in tissue mechanics but have not been applied to GBM \citep{Osborne2010, Alt2017}.

Table~\ref{tab:DMs} gives an overview of all discrete methods discussed above.

\begin{table}
    \centering
    \caption{Overview of the reviewed discrete models. The references are not exhaustive. For other recent reviews, see \citep{Liedekerke2015, Wang2015, Harris2019, Metzcar2019, Weerasinghe2019}.}
    \vspace{0.1cm}
    \label{tab:DMs}
    \rowcolors{2}{orange!10}{white}
    \begin{tabular}{|p{0.18\textwidth}|p{0.42\textwidth}|p{0.34\textwidth}|}
        \hline
        \rowcolor{orange!20} \bf Model type & \bf Features & \bf References \\
        \hline
        Cellular automata \newline (CA) models & \textbf{Concept:} All CA-type models are discrete and lattice-based. Cell migration, death, movement, division, and growth are modelled as stochastic processes. \textbf{Particularity:} Multiple cells or at most one are allowed per lattice site. \newline \textbf{Assets:} Centimetre-scale simulations are feasible when tracking cell clusters. Individual cells can be represented.
        \newline \textbf{Drawbacks:} They do not represent cell morphology, have a limited spatial resolution and are computationally heavy. For multiple cells per site, the individual cell positions are not tracked. & \textbf{One cell per lattice site:} \citep{Radszuweit2009, Jiao2013EvolutionAM, Jagiella2016, Poleszczuk2016}
        \newline
        \textbf{Multiple cells per lattice site:} \citep{Kansal2000, Schmitz2002}
        \newline
        \textbf{Bridging scales:} Varying number of cell limits per site at different locations \citep{Liedekerke2015}.
        \\ [2pt]
        Lattice gas \newline cellular automata \newline (LGCA) models & \textbf{Concept:} They are CA-type models. There are multiple velocity channels (and hence cells) per lattice site. A deterministic step evaluating movement is included (see Fig.~\ref{fig:LGCA}).
        \newline \textbf{Assets:} Individual cells are represented. Convergence to CM has been shown.
        \newline \textbf{Drawbacks:} They do not represent cell morphology, have a limited spatial resolution and are computationally heavy. & 
        Deutsch, Hatzikirou and collaborators have investigated cell proliferation and glioma invasion \citep{Chopard2010, Hatzikirou2010, Hatzikirou2012, Tektonidis2011}.
        \\ [2pt]
        Cellular Potts \newline models (CPM) & \textbf{Concept:} They are CA-type models. There are multiple sites per cell. A Hamiltonian function governs stochastic events. 
        \newline \textbf{Assets:} CPMs draw on the physics behind the cell dynamics and represent cell morphology.
        \newline \textbf{Drawbacks:} They have a limited spatial resolution. Centimetre-scale simulations are not feasible. & 
        Authors have included a broad range of phenomena through a variety of terms in the Hamiltonian \citep{Ouchi2003ImprovingTR, Shirinifard2009, Gao2012, Daub2013, Szabo2013}.
        \\ [2pt]    
        Centre-based \newline models (CBM) & \textbf{Concept:} CBMs model cells as physical particles governed by their equations of motion. No lattice is used. Cells are simple geometrical objects.  
        \newline \textbf{Assets:} The physics behind the cell dynamics is accounted for. Cells and their movement are resolved.
        \newline \textbf{Drawbacks:} CBMs are computationally heavy. Centimetre-scale simulations are not feasible. Cell morphology is not modelled. & 
        Cells can be spherical \citep{Drasdo2005, Hatzikirou2010} or ellipsoidal \citep{Dallon2004HowCM}. Studies of tumour cell drug response: \citep{Perez-Velazquez2020}.
        \\ [2pt] 
        Deformable cell \newline models (DCM) & 
        \textbf{Concept:} Cells are depicted as a collection of connected nodes described by their equations of motion.
        \newline \textbf{Assets:} DCMs model cell morphology and stresses on a subcellular level. There is no lower length scale since no lattice is used.
        \newline \textbf{Drawbacks:} Only simulations of small cell populations or single cells are feasible.
        & Single cell adhesion and response to mechanical stress \citep{Liedekerke2010, Odenthal2013}. Tumour cell proliferation \citep{Rejniak2007}. Vertex models \citep{Alt2017}. 
        \\ [2pt]
        \hline
\end{tabular}
\end{table}


\section{Hybrid multi-scale models} \label{sec:hybrid}

In the literature, the term hybrid or multi-scale model might denote any combination of the different methods discussed above. A large variety of approaches thus fall into this category. Adopting the nomenclature from Deisboeck et al. \citep{Deisboeck2010}, we distinguish between three types: \textit{composite} hybrid models, \textit{adaptive} hybrid models, and calibrated models, such as continuum models with functional parameters. Composite hybrid models track and connect phenomena across various scales while keeping the scale associated with each individual aspect fixed (cf. Section~\ref{sec:composite}). Thus, all tumour cells are treated on equal footing.

In contrast, adaptive hybrid models dynamically adjust the local resolution based on the required level of detail (cf. Section \ref{sec:adaptive}). That is to say that not all tumour cells are represented in the same manner. Finally, the third type of hybrid model denotes, for instance, continuum models, whose parameters have been calibrated based on agent-based models or biophysical considerations of microscopic or mesoscopic phenomena (cf. Section~\ref{sec:funcparam}).


\subsection{Composite hybrid models} \label{sec:composite}

When using a discrete model to represent tumour cells, the extracellular biochemical players, such as oxygen, are commonly modelled as continuous fluids~\citep{powathil2012modelling, powathil2015systems, powathil2013towards, kim2018role}. Moreover, intracellular changes, i.e. processes on a sub-cellular scale, might be tracked for each individual cell using ODEs as discussed in Section~\ref{sec:ODE}. These different microscopic, mesoscopic, and macroscopic mechanisms are all coupled. For instance, the nutrient levels both depend on the cell density and impact the probabilities that are used to predict cell actions. When this interdependence is taken into account, the discrete tumour cell model becomes one of many components in a larger framework that spans and links multiple temporal and spatial scales. The framework as a whole then constitutes a composite hybrid model \citep[cf.][]{Athale2005, Anderson2006, Athale2006, Wang2007, Kim2011, Kim2013, Kim2014, Caiazzo2015, Szabo2017, Antonopoulos2019, Chamseddine2019, Gallaher2020}. Such hybrid frameworks have been used to study a wide variety of phenomena, including drug resistance \citep[e.g.][]{Engblom2018, Perez-Velazquez2020}. In each case, the focus of the study prescribes the resolution of different features. For instance, in some studies, discrete models of the vasculature are combined with continuum models of the tumour to better understand angiogenesis or drug delivery \citep[e.g.][]{Chamseddine2018}. As another example, May et al. \citep{May2011} couple discrete tumour models with biomechanical calculations of stresses and strains to arrive at a more accurate prediction of the tumour shape. Finally, we note that composite hybrid models have been employed in Bayesian inferences to establish posterior probability distributions for relevant model parameters \citep{Jagiella2017} (see also Section~\ref{sec:data}).


\subsubsection{Ordinary differential equations} \label{sec:ODE}

When dealing with agent-based hybrid models, some authors include sub-cellular dynamics, modelling molecular networks, pathways, and reactions \citep{Athale2005, Athale2006, Zhang2009, KimY2015}. Such processes are governed by a system of coupled nonlinear Ordinary differential equations (ODEs)  that represent mass-balance equations and rely on the stoichiometry of the underlying biochemical reactions.

ODEs thus play a vital role in some composite hybrid models. In general, ODEs are a common tool in mathematical biology and oncology. For instance, they can capture global tumour growth patterns \citep{Sachs2001SimpleOM}. While such models do not spatially resolve the tumour, they have successfully been used to study the response to drugs and radiation treatments, recovering experimental constraints on the overall growth pattern \cite[e.g.][]{Simeoni2004, Koziol2020}. The ODE for tumour growth take the generic form
\begin{equation}
\frac{\mathrm{d} N(t)}{\mathrm{d} t} = S(N)
\end{equation}
where $N(t)$ is the total cell number or volume, and $S(N)$ denotes an appropriate source function.


\subsection{Adaptive hybrid models} \label{sec:adaptive}

Detailed modelling of individual cells is required to capture tumour growth at low cell numbers. Agent-based models thus give unique insights into glioma invasion. However, at high cell densities, a lower resolution suffices. Continuum models realistically capture all relevant aspects of large chunks of bulk tumours. Adaptive hybrid models take advantage of this notion by describing different parts of the tumour tissue using distinct modelling approaches. Thereby, it becomes possible to draw on the strengths of discrete models when depicting centimetre-sized tumours without the insurmountable computational cost that would otherwise be involved.

The hybrid model presented by the group of Stolarska and Yangjin \citep{Kim2007, Stolarska2009, KimY2015} treats necrotic tumour zones, quiescent tumour tissue, and the surrounding microenvironment as continuous fluids. The relevant PDEs are solved on a regular grid. The cells in the proliferative tumour rim (100-\SI{200}{\micro\metre}), on the other hand, are modelled using a CBM. They are represented as autonomous ellipsoids. At the boundary between the cell-based and continuous components, interpolation is used to establish the boundary conditions for the reaction-diffusion equations and the forces exerted on the individual agents \citep[see also][for a related example from molecular dynamics]{Erban2016}.

Bearer et al. \citep{Bearer2009} and Frieboes et al. \citep{Frieboes2010} couple an off-lattice approach with the continuum models by Wise et al. \citep{Wise2008} discussed in Section~\ref{sec:lineage}. To study glioma invasion, they include dynamic transitions between the continuum and cell-based representations based on the microenvironment. In their model, both mutations and hypoxia might induce the development of a migratory phenotype \citep[see also][]{Rocha2021}.

The adaptive hybrid models discussed above allow for mass transfer between the different components and are carefully constructed to obey laws, such as mass conservation. However, it is worth noting that the continuum models are pre-assumed without guaranteeing that the imposed functional form emerges from up-scaling the cell-based representation (cf Section~\ref{sec:funcparam}). Moreover, due to the assumptions that enter the continuum and agent-based models, single-cell measurements might lead to different parameter values than those required for continuum models to fit patient-specific data \citep{Klank2018}.

Rather than coupling agent-based and continuum models, other authors \citep{Zhang2009b, Zhang2011, Liedekerke2015} combine agent-based models with different resolutions. Such approaches are known as multi-scale agent-based modelling. For instance, one might couple CA models that consider single cells with models that deal with cell clusters at a coarser lattice resolution \citep[see also][for a multi-resolution study of polymers]{Rolls2017}.


\subsection{Calibrated models} \label{sec:funcparam}

Various authors encode information from cell-based models or biophysical considerations into heterogeneous and time-varying parameters of continuum models. This goal can be achieved in different ways. As briefly discussed in Section~\ref{sec:LGCA}, one might up-scale agent-based models using the mean-field approximation \citep[cf.][]{Hatzikirou2010}. Other authors draw on the emergent macroscopic properties of agent-based models to derive phenomenological relationships \citep[cf.][]{Lowengrub2010, Macklin2012}. 

Each method mentioned above relies on simplifying approximations to derive an analytical expression, which might limit their predictive power. To circumvent the need to explicitly state expressions for the macroscopic variables altogether, Kavousanakis et al. \citep{Kavousanakis2012} use a form of equation-free modelling called coarse projective integration \citep[see also][]{Aviziotis2015}. This method invokes the so-called coarse time-stepper that consists of four phases: lifting, simulation, restriction, and extrapolation. These four phases are repeated for every macroscopic time step. In the lifting step, an ensemble of cell distributions is drawn from the macroscopic tumour cell density. Each cell distribution is then evolved over a short time span using an agent-based model --- Kavousanakis et al. use a CA model. In the restriction step, the macroscopic density is updated by computing the ensemble average of the agent-based simulations. Under the assumption that the macroscopic cell density evolves more slowly than the microscopic variables, the observed change in the cell density yields an estimate of its time derivative. Using the forward Euler method, this information can be employed to project the cell density further into the future without considering individual cells.

We note that the general idea of calibrating coarse models based on more resolved representations is not only limited to continuum models. Recently, Van Liedekerke et al. \citep{Liedekerke2019} have thus calibrated the mechanical interaction forces of CBMs based on DCMs.


\section{Data-driven modelling} \label{sec:data}

Researchers draw on a wide range of experimental data to inform their mathematical models of GBM and do so following several different approaches. One can use data to select the underlying mechanism or inform the model parameters. For example, as discussed above, some researchers use DTI MRI data to determine the diffusion properties of cancer cells. Moreover, data allow for qualitative comparisons and constraints on the simulations. For instance, most models are set to recover generic properties of cancer, such as spherical avascular tumour growth. Finally, one can use data to infer model parameter values using a statistical framework, either maximum likelihood estimation or a Bayesian approach. Systematic incorporation of imaging and molecular data into the mathematical models of cancer via statistical inference is a promising and timely area of future development. This area of development is bolstered by the latest progress in imaging and omics approaches which are producing quantitative data on an unprecedented scale.
Furthermore, the latest progress in computational power, simulation-based inference methods, and machine learning make this task computationally feasible. Several modelling studies aim to fit specific data from patients, animal models, or \textit{in vitro} experiments. Here, we provide a list of illustrative examples and summarise our perspective on important areas of future research. We also refer the reader to some recent reviews on the topic \citep{Falco2021, Martinez2022}. 

Two decades ago, Stamatakos and collaborators \citep{Stamatakos2006, Stamatakos2006b} used agent-based models to investigate the impact of chemotherapy \textit{in vivo} based on patient-specific PET, SPECT, T1-weighted MRI, histopathologic and genetic data. They reduced the computational cost by clustering the cells into macroscopic (1$\,$mm$^2$) regions rather than simulating individual cells. More recently, Gallaher et al. \citep{Gallaher2020} published a study in which they fit MRI and \textit{ex vivo} cell tracking data from rats to off-lattice agent-based models, simulating individual cells. They accomplish this fit by using random sampling. However, due to the high computational cost, agent-based models are most commonly found in studies focusing on small cell populations. For instance, Oraiopoulou et al. \citep{Oraiopoulou2018} use agent-based models to recreate the invasive morphologies of GBM observed \textit{in vitro}. Meanwhile, many papers on agent-based models do not include a fit to real-world data but rather perform \textit{in silico} experiments to make qualitative and quantitative predictions. For instance, based on such analyses, Perez-Velazquez et al. \citep{Perez-Velazquez2020} addressed the development of drug resistance, Kim and collaborators \citep{Kim2013, Kim2014} investigated the key mechanisms behind molecular switches, and Schmitz et al. studied growth patterns.

The parameters used in state-of-the-art continuum models \citep{Klank2018} to capture the migration and proliferation of GBM are derived from MRI and CT images, and the application of continuum models to patient-specific data \textit{in vivo} is widespread in the literature. Thus, Swanson et al. \citep{Swanson2000, Swanson2002} already used such data to inform their models twenty years ago \citep[see also][]{Harpold2007}. Later studies that build on their work have included further imaging data, such as PET \citep{WangRockhill2009, Gu2012, Neal2013}. Isotropic diffusion reaction equations yield a detailed picture based on T1Gd and T2 Flair MRI images \citep{Corwin2013}. Other studies have employed histopathology \citep{Frieboes2007}. The anisotropic continuum models discussed in Section~\ref{sec:anidiff} likewise built on MRI data, drawing on DTI images to compute the diffusion matrix \citep{Jbabdi2005, PainterHillen2013, Engwer2015, Angeli2018, Swan2018}. 

Due to the relatively low computational cost of continuum models, they can be employed in Bayesian sampling schemes as illustrated by Lipkova et al. \citep{Lipkova2019}, who infer model parameters based on MRI and PET images using a Markov Chain Monte Carlo algorithm. Lipkova et al. demonstrate that their Bayesian approach can be used to improve personalised radiotherapy. Ezhov et al. \citep{2021arXiv211104090E} address the same inverse problem using machine learning. Ezhov et al. construct a library of 100,000 simulations with different parameter values and use their atlas to train a neural network. The trained neural network can infer the patient-specific model parameters based on MRI images within minutes, allowing for effortless model personalisation. Today, continuum models are thus at the point where they have the potential to enter clinical settings and contribute to personalised treatments. While the same does not yet hold for agent-based models due to their high computational cost, machine learning might likewise help to overcome this obstacle in the future, as discussed by Joergensen et al. \citep{Joergensen2021}. Machine learning algorithms would play the same role as they do in continuum models by circumventing the need to produce further simulations once a surrogate model or inference algorithm has been trained. Similar strategies are applied when dealing with agent-based models outside of cancer research \citep{PMID:34893600}.

Like imaging data, molecular data offers unique insights into GBM \citep{Brennan2013, McLendon2008}. In particular, different sequencing techniques, including bulk RNA-seq \citep{brooks2021white}, scRNA-seq \citep{neftel2019integrative, bhaduri2020outer, couturier2020single, White2020, xie2021key, richards2021gradient} (recently reviewed in \citep{cells10092264} and \citep{Karaayvaz2018}) and spatial RNA-seq \citep{ravi2022spatially}, have helped researchers to uncover different aspects of brain tumours. Sequencing data have thus contributed to our understanding of important cell types, the nature of the invasive tumour front, the spatial cellular organisation of the tumours, and the molecular basis of the heterogeneity of GBM. For instance, Ravi et al. \citep{ravi2022spatially} recently identified spatially distinct clusters of genes via spatial transcriptomics, and Neftel et al. \citep{neftel2019integrative} identified four main cellular states among malignant tumour cells by analysing data at the single-cell level. Furthermore, Neftel et al. demonstrated that the relative frequency of these four states varies between GBM samples. On the one hand, such information on the heterogeneity of GBM is valuable in its own right and can help to inform the underlying assumptions used in the mathematical models. As exemplified in Section~\ref{sec:lineage}, mathematical models thus present a unique tool to investigate the implications of such information on the tumour phenotype. On the other hand, the vast number of spatial GBM sequencing data that is becoming available can directly be used in parameter inference, model selection or validation of model predictions.


\section{Clinical application of spatio-temporal modelling in Glioblastoma} \label{sec:Clinical_app}

Glioblastoma is a biologically complex and dynamic tumour that exists within the intricate and responsive brain environment. Parcelling and modelling discreet aspects of these biological processes and systems holds the potential for numerous direct clinical care applications and has the potential to address several areas of unmet need.

Mathematical diffusion-tensor modelling is currently widely used in the context of MRI diffusion-tensor-based tractography for the modelling of white-matter tracts. Understanding patient-specific anatomy, and anatomico-pathological relationships, is critical for effective surgical planning and avoidance of iatrogenic injury \citep{PMID:32176926}. Future refinements in data modelling can be expected to improve spatial resolution and reliability, which would directly translate to improved validity of neuro-imaging with direct clinical applications (cf. Section~\ref{sec:data}). 

A key concern in clinical practice is the prediction of tumour growth patterns, sites of progression, and the location of tumour recurrence. This triad is important both for therapeutic planning and prognostication \citep{PMID:33251147}. Mathematical models have the potential to refine these facets of care and hence to offer highly-tailored conformal radiation dosing and targeted surgical supra-marginal resection \citep{PMID:34098965, PMID:34087795}. These therapeutic adaptations might maximise the prognostic benefits and lead to the destruction of the tumour while sparing normal neural tissue. Similarly, when combined with the modelling of growth rates, this information could be assimilated to generate more accurate prognostic maps with the prediction of anatomico-pathological functional deficits and overall prognosis. 

Recent years have been characterised by increasing recognition of glioblastoma's molecular intra- and inter-tumoural heterogeneity and its functional consequences. This is exemplified in the evolution of the WHO classification of tumours from a purely histopathological to a fully-integrated molecular classification \citep{PMID:34185076}. As we have summarised, mathematical models of glioblastoma hold the potential to illustrate novel prognostic aspects of tumour biology and the micro-environment not currently captured. These might include structural metrics, such as tissue density and cellular linage, or functional ones, including oxygenation, nutrient supply and metabolism. 

The potential for applying spatio-temporal mathematical modelling of glioblastoma in clinical care is substantial but thus far remains largely unrealised. As summarised in this review, the foundation of knowledge in this area is now considerable. The stage is set for those who wish to advance the field from theoretical models to practical applications.


\section{Summary and discussion} \label{sec:disc}

This paper presents an overview of mathematical models that provide means to scrutinise and predict the spatial and temporal evolution of tumours. While we focus on models for glioblastoma and highlight characteristics of this aggressive cancer type, we also attempt to give a rounded picture of state-of-the-art approaches across the field of mathematical oncology.

Spatially resolved tumour models fall into two main categories: They either treat the tumour cell density as a continuous variable (cf. Section~\ref{sec:continuum}) or deal with autonomous agents representing individual cells or cell clusters (cf. Section~\ref{sec:discrete}). 

The computational cost associated with continuous models is relatively low. This property allows for the modelling of the entire tumour volume and the application to a wide variety of (patient-specific) data. Furthermore, continuous models can capture many of the characteristics of GBM. We summarise how continuous models achieve this goal in Table~\ref{tab:CMbsAGB}. GBM spreads anisotropically following existing structures in the brain, including white matter tracks and blood vessels. On a macroscopic level, this behaviour is well-described by reaction-diffusion equations when including an anisotropic diffusion term (cf. Section~\ref{sec:anidiff}).

Moreover, by including source terms that couple a set of such reaction-diffusion equations, continuous models can link different cell sub-populations and account for interactions with the environment (cf. Section~\ref{sec:microenv}). Continuous models are hereby able to encapsulate the hierarchical nature that is a defining feature of GBM (cf. Section~\ref{sec:lineage}). The models also show great versatility: By including additional biologically motivated terms into the reaction-diffusion equations, continuous models can be painlessly extended to include phenomena ranging from immune responses to the impact of radiotherapy.

\begin{table}
    \centering
    \caption{Overview of how different model types can deal with properties that define glioblastoma growth.}
    \vspace{0.1cm}
    \label{tab:CMbsAGB}
    \rowcolors{2}{orange!10}{white}
    \begin{tabular}{|p{0.18\textwidth}|p{0.38\textwidth}|p{0.38\textwidth}|}
        \hline
        \rowcolor{orange!20} \bf Phenomenon & \bf Continuum models & \bf Agent-based models \\
        \hline
        Invasion of healthy tissue & Diffusion (and advection) terms are included in the PDE. & Stochastic (and deterministic) rules dictate the migration of individual cells.
        \\ [2pt]
        Anisotropies in tumour growth & E.g. anisotropic diffusion tensor is used based on brain structure. & By allowing for feedback between the cells and environment, one can introduce preferential migration, leading to anisotropic growth along structures.
        \\ [2pt]
        Cell hierarchy and lineage & A set of coupled PDEs described subpopulations with different properties. & Individual cells underlie rules that are affected by cell type.
        \\ [2pt]
        Mitosis and apoptosis & E.g. logarithmic or exponential source terms are included in the PDE. These might link different sub-populations and be associated with properties of the environment. & Both phenomena underlie stochastic or deterministic rules that are invoked for each agent.
        \\ [2pt]
        \hline
\end{tabular}
\end{table}

Tumour cells must always be understood in relation to their surroundings. In this connection, aspects, such as nutrition flows, the extracellular matrix, or the vasculature, are often (if not always) represented using continuous models. Even when using agent-based models, continuous models thus play a vital role. However, continuous models have their drawbacks. They do not account for the stochastic events that dominate at low tumour cell densities. Consequently, they are not ideal for addressing some questions regarding the invasive tumour front or tumour recurrence that is a prevalent obstacle in treating GBM.

At low cell densities, agent-based models have an edge. Such models come in many different flavours. Some agent-based models borrow from the description of other phenomena. They thus lean on theoretical models, e.g. from solid-state physics, posing biological events in terms of physical forces and potentials (cf. Sections~\ref{sec:CPM}, \ref{sec:CMs}, and \ref{sec:DMs}). Other models rely on purely stochastic rules to encode biological information (cf. Sections~\ref{sec:CA} and \ref{sec:LGCA}). Moreover, agent-based models differ in the details that they encompass. While some models address aspects of cell morphology, others do not. While some models define a lower threshold for their spatial resolution by employing a lattice on which the agents move, others do not.

Like continuous models, agent-based models are able to capture the essential properties of GBM (see Table~\ref{tab:CMbsAGB}). Anisotropic diffusion along existing structures can be included by altering the rules that govern cell migration based on the environment. Moreover, by altering these rules based on the cell type, the hierarchical organisation of GBM can be incorporated into the model.

The high computational expense of agent-based models constitutes their main drawback. It can be a significant stumbling block on the way to unleashing their full potential. However, this shortcoming can be partly circumvented. One might thus combine different approaches into a hybrid model bridging across the scales involved (cf. Sections~\ref{sec:composite} and \ref{sec:adaptive}). Alternatively, one might study small cell populations to derive emergent macroscopic properties that can be used to calibrate models on a coarser scale (cf. Section~\ref{sec:funcparam}). Indeed, one of the main assets of agent-based models is their ability to link an understanding of individual cells to macroscopic events. To further lower the computational cost when confronting the models with data, one might use sophisticated statistical techniques or machine learning methods that demonstrably reduce the number of models that need to be constructed \citep[see also][]{2021arXiv210104653L, Joergensen2021}, as discussed in Section~\ref{sec:data}.

Each of the available mathematical models has its strength and weaknesses. Which model to employ will hence be dictated by the research question and the available data. First, some approaches are better suited for depicting certain phenomena than other methods are. One might drastically simplify the work associated with the practical implementation by choosing the method wisely. Secondly, not all models are equally informative. The question is whether the additional information can be exploited and for what purpose \citep[see also][]{Enderling2020}. On the one hand, the chosen model sets limits on the insights that one might gain. This notion favours more complex models. On the other hand, one should avoid cracking nuts with a sledgehammer. To exemplify this, it is thus worth noting that \textit{non-spatial models} make full use of certain types of (patient-specific) data whilst being unable to address any questions on tumour morphology. They are undoubtedly very valuable tools in their own right. The same holds for each spatio-temporal cancer model presented above. With this in mind, we hope that our review provides a useful guide and helps the reader select a suitable mathematical model for their research.

Mathematical models are only useful if they describe and relate to experimental data. Our review thus ends with a summary and discussion of the application of mathematical models to experimental data and its clinical relevance. A wealth of imaging and molecular data is becoming available, particularly due to advances in single-cell and spatial transcriptomics. At the same time, researchers have begun to implement spatial models into statistical frameworks to perform systematic inference based on these data. While few such inference analyses have yet been published, advances in machine learning and computing power promise to further enable the systematic incorporation of imaging and molecular data into the spatial models of GBM. We thus envision a greater impact of mathematical modelling in unravelling the basic biology of GBM and its applications in clinical treatment in the near future.


\section*{Acknowledgement}
This work was supported by The Oli Hilsdon Foundation through The Brain Tumour Charity, grant number (GN-000595), in connection with the program ``Mapping the spatio-temporal heterogeneity of glioblastoma invasion". Work at the Cancer Institute Genomics Translational Technology Platform is supported by the CRUK City of London Centre Award [C7893/A26233].


\bibliographystyle{acm}

\end{document}